\documentclass{article}
\usepackage{ltwol2e}
\usepackage{epsfig}
\font\eightrm=cmr8
\begin{document}
\twocolumn[
\begin{flushright}
DESY 98-128\\
September 1998\\
\end{flushright}
\vspace*{1.5cm}
\begin{center}
{\Large \bf
\centerline{TOWARDS A THEORY OF CHARMLESS NON-LEPTONIC}
\vspace*{0.2cm}
\centerline{ TWO-BODY $B$ DECAYS}}  
 \vspace*{1.5cm}
%\vskip1cm
 {\large A.~Ali}
\vskip0.2cm
 Deutsches Elektronen-Synchrotron DESY, Hamburg \\
Notkestra\ss e 85, D-22603 Hamburg, FRG\\  

\vspace*{8.0cm}
{\large
Invited Talk; To be published in the Proceedings of the
XXIX International Conference\\ on High Energy Physics, 
Vancouver, B.C., Canada,  July 23 - 29, 1998}

\end{center}]

\newpage
\font\eightrm=cmr8

\arraycolsep1.5pt 

% A useful Journal macro
\def\Journal#1#2#3#4{{#1} {\bf #2}, #3 (#4)}

% Some useful journal names
\def\NCA{\em Nuovo Cimento}
\def\NIM{\em Nucl. Instrum. Methods}
\def\NIMA{{\em Nucl. Instrum. Methods} A}
\def\NPB{{\em Nucl. Phys.} B}
\def\PLB{{\em Phys. Lett.}  B}
\def\PRL{\em Phys. Rev. Lett.}
\def\PRD{{\em Phys. Rev.} D}
\def\ZPC{{\em Z. Phys.} C}
% Some other macros used in the sample text
\def\st{\scriptstyle}
\def\sst{\scriptscriptstyle}
\def\mco{\multicolumn}
\def\epp{\epsilon^{\prime}}
\def\vep{\varepsilon}
\def\ra{\rightarrow}
\def\ppg{\pi^+\pi^-\gamma}
\def\vp{{\bf p}}
\def\ko{K^0}
\def\kb{\bar{K^0}}
\def\al{\alpha}
\def\ab{\bar{\alpha}}
\def\be{\begin{equation}}
\def\ee{\end{equation}}
\def\bea{\begin{eqnarray}}
\def\eea{\end{eqnarray}}
\def\CPbar{\hbox{{\rm CP}\hskip-1.80em{/}}}%temp replacemt due to no font

\bibliographystyle{unsrt}    %for BibTeX - sorted numerical labels

\newcommand{\dis}{\displaystyle}
\newcommand{\lsim}{\stackrel{<}{_\sim}}
\newcommand{\gsim}{\stackrel{>}{_\sim}}
\def\s{\hat{s}}
\def\u{\hat{u}}
\def\z{v \cdot \hat{q}}
\def\cnt{C_9^{\mbox{eff}} \mp C_{10}}
\def\ml{\hat{m}_l}
\def\ms{\hat{m}_s}
\def\mx{\hat{m}_x}
\def\mc{\hat{m}_c}
\def\mb{\frac{M_B}{{m_b}^3}}
\def\lo{\hat{\lambda}_1}
\def\lt{\hat{\lambda}_2}
\def\loo{\lambda_1}
\def\lto{\lambda_2}
\def\q{\hat{q}}
\def\bxsll{$B \rightarrow X_s \ell^+ \ell^- $}
\def\absvcb{\left| V_{cb} \right|}

% for citer
\catcode`@=11
% --------------------------

% Original Latex definition of citex, except for the removal of
% 'space' following a ','

\def\@citex[#1]#2{\if@filesw\immediate\write\@auxout{\string\citation{#2}}\fi
  \def\@citea{}\@cite{\@for\@citeb:=#2\do
    {\@citea\def\@citea{,\penalty\@m}\@ifundefined
      {b@\@citeb}{{\bf ?}\@warning
       {Citation `\@citeb' on page \thepage \space undefined}}%
\hbox{\csname b@\@citeb\endcsname}}}{#1}}

\def\citer{\@ifnextchar [{\@tempswatrue\@citexr}{\@tempswafalse\@citexr[]}}

% \citer as abbreviation for 'citerange' replaces the ',' by a '--'
%

\def\@citexr[#1]#2{\if@filesw\immediate\write\@auxout{\string\citation{#2}}\fi
  \def\@citea{}\@cite{\@for\@citeb:=#2\do
    {\@citea\def\@citea{--\penalty\@m}\@ifundefined
       {b@\@citeb}{{\bf ?}\@warning
       {Citation `\@citeb' on page \thepage \space undefined}}%
\hbox{\csname b@\@citeb\endcsname}}}{#1}}
% --------------------------
\newcommand{\postbb}[3]
{\setlength{\epsfxsize}{#3\hsize}
\centerline{\epsfbox[#1]{#2}}}
\newcommand{\optbar}[1]{\shortstack{{\tiny (\rule[.4ex]{1em}{.1mm})}
  \\ [-.7ex] $#1$}}
%%%%%%%%%%%%%%%%%%%%%%%  NEWCOMMANDS %%%%%%%%%%%%%%%%%%%%%%%%%%%%
%\renewcommand{\textfraction}{0.0}
%\renewcommand{\topfraction}{1.0}
%\renewcommand{\bottomfraction}{1.0}
%   DINA4 format DESY
% end dina4 format from DESY
%\renewcommand{\theequation}{\thesection.\arabic{equation}}
\newcommand{\epsl}{\varepsilon \hspace{-5pt} / }
\newcommand{\rsl}{r \hspace{-5pt} / }
\newcommand{\qsl}{q \hspace{-5pt} / }
\newcommand{\pbsl}{p \hspace{-5pt} / }
\newcommand{\pssl}{p' \hspace{-7pt} / }
\newcommand{\hm}{\hat{m}_c^2}
\newcommand{\BR}{\mbox{BR}}
\def\vub{V_{ub}}
\def\vcb{V_{cb}}
\def\vtb{V_{tb}}
\def\vuqst{V_{uq}^*}
\def\vcqst{V_{cq}^*}
\def\vtqst{V_{tq}^*}
\def\qsqav{<q^2>}
\newcommand{\ba}{\begin{array}}
\newcommand{\ea}{\end{array}}

% Bars on quarks:
\def\qb{\bar{q}}
\def\ub{\bar{u}}
\def\db{\bar{d}}
\def\cb{\bar{c}}
\def\sb{\bar{s}}

% Bra-Kets:
\def\bra{\langle}
\def\ket{\rangle}

% Greek letters:
\def\a{\alpha}
\def\b{\beta}
\def\g{\gamma}
\def\d{\delta}
\def\e{\epsilon}
\def\p{\pi}
\def\ve{\varepsilon}
\def\ep{\varepsilon}
\def\et{\eta}
\def\l{\lambda}
\def\m{\mu}
\def\n{\nu}
\def\G{\Gamma}
\def\D{\Delta}
\def\L{\Lambda}
% Specials:
\def\to{\rightarrow}
%%%%%%%%%%%%%%%%%%%%%%%%%%%%%%%%%%%%%%%%%%%%%%%%%%%%%%%%%%%%%%%%%%%%%%%%%
%                      BEGINNING OF TEXT
%%%%%%%%%%%%%%%%%%%%%%%%%%%%%%%%%%%%%%%%%%%%%%%%%%%%%%%%%%%%%%%%%%%%%%%%%
       
\title{TOWARDS A THEORY OF CHARMLESS NON-LEPTONIC TWO-BODY $B$ DECAYS}

\author{A. ALI}
 
\address{Deutsches Elektronen-Synchrotron DESY,
Hamburg, Germany\\E-mail: ali@x4u2.desy.de}


\twocolumn[\maketitle\abstracts{We address a number of theoretical 
and phenomenological issues in two-body charmless
non-leptonic $B$ decays in the context of a factorization model. A
classification of the exclusive decays involving tree and penguin amplitudes
is reviewed. The role of QCD anomaly
in exclusive decays involving an $\eta$ or $\eta^\prime$ is elucidated and  
comparison is made with the existing data. We argue that the factorization
approach accounts for most of the observed two-body $B$ decays.}]

\section{Introduction}
The standard theoretical framework to study $B$ decays is
based on an effective Hamiltonian, obtained by integrating out the
top quark and $W^\pm$ fields,
which allows to separate the short- and long-distance-contributions
in these decays using the operator product expansion.
QCD perturbation theory is used in
deriving the renormalization group improved short-distance
contributions and in evaluating the matrix elements at the parton level.
The long-distance part in the two-body hadronic decays $B \to M_1 M_2$
involves  the transition matrix elements $\langle M_1 M_2 |
O_i |B \rangle$, where $O_i$ is a four-quark or magnetic moment operator. 
Calculating these matrix elements from  first principle is a true challenge 
and a quantitative theory of exclusive $B$ 
decays is not yet at hand. Hence,  some  assumption about  
handling the hadronic matrix elements is at present
unavoidable. We assume factorization, in which soft final state 
interactions (FSI) are ignored, and hence the hadronic matrix 
elements in the decays $B \to M_1 M_2$ factorize into a product of  
theoretically more tractable quantities \cite{BSW87}.

The rationale of factorization lies in the phenomenon of 
colour-transparency \cite{Bjorken} in which 
a pair of fast moving (energetic) quarks in a colour-singlet state 
effectively decouples from long-wavelength gluons. In two-body $B$-decays, 
the decay products have each an energy $E_i$ of $O(m_B/2)$, which is large 
enough (compared to $\Lambda_{QCD}$) for the above argument to hold.
Phenomenologically, the factorization framework does remarkably 
well in accounting for the observed non-leptonic two-body $B$-decays 
involving the
current-current operators $O_{1,2}^{c}$, inducing $b \to c$ transitions 
\cite{NS97}. The decays $B \to h_1 h_2$, where
$h_1$ and $h_2$ are light hadrons, are more complex as they involve, apart 
from the current-current, also QCD- and electroweak 
penguin-induced amplitudes.
 However, in simpler circumstances where a single (Tree 
or Penguin) amplitude
dominates, it should be possible to make predictions for the decays $B 
\to h_1 h_2$ in the factorization framework which are on the same
theoretical footing as the two-body $B$-decays governed by the operators 
$O_{1,2}^{c}$. We review  
some selected $B \to h_1 h_2$ decays. The underlying theoretical framework 
and the results presented here are 
based on the work done in collaboration
with Greub \cite{AG97}, Chay, Greub and Ko \cite{ACGK97}, and 
Kramer and L\"u \cite{AKL98-1,AKL98-2}. A comparison with the available data
from the CLEO collaboration, some of which has been updated at this 
conference \cite{cleo}, is also made.
%%%%%%%%%%%%%%%%%%%%%%%%%%%%%%%%%%%%%%%%%%%%%%%%%%%%%%%%%%%%%%%%%%%%%%%%%%%%%
\section{Effective Hamiltonian Approach to $B$ Decays}
The effective Hamiltonian for the $\Delta B=1$ transitions can be written as:
\bea
H_{eff}
&=& \frac{G_{F}}{\sqrt{2}} \, [ \vub \vuqst
\left (C_1 O_1^u + C_2 O_2^u \right) \nonumber\\  
{} &+& \vcb \vcqst
\left (C_1 O_1^c + C_2 O_2^c \right) -
\vtb \vtqst \, 
\sum_{i=3}^{12}
C_{i} \, O_i ],
\eea
where $G_F$ is the Fermi coupling constant, $V_{ij}$ are the CKM 
matrix elements, $C_i$ are the Wilson coefficients and $q=d,s$. The 
counting of these operators is as follows:
$O_{1,2}^u$ and $O_{1,2}^c$ are the current-current (Tree) operators which
induce the $b \to u$ and $b \to c$ transitions, respectively; 
$O_{3},...,O_{6}$ are the
QCD-penguin operators, $O_{7},...,O_{10}$ are the Electroweak-penguin 
operators, and $O_{11} (O_{12})$ is the electromagnetic (chromo-magnetic)
dipole operator. Their precise definition can be seen 
elsewhere \cite{AKL98-1}. The Wilson coefficients depend (in general) on 
the renormalization 
scheme and the scale $\mu$ at which they are evaluated. However, the
physical matrix elements $\langle h_1 h_2 | H_{eff} | B \rangle$ are
obviously independent of both the scheme and the scale. Hence, the 
dependencies in the Wilson coefficients must 
be compensated by a commensurate calculation of the hadronic matrix 
elements in a non-perturbative framework, such as 
lattice QCD. Presently, this is not a viable strategy 
as the calculation of the matrix elements 
 $\langle h_1 h_2 | O_i|B \rangle$ is beyond the scope
of the current lattice technology. However, perturbation theory comes to 
(partial) rescue, with the help of which one-loop matrix
elements can be rewritten in terms of the tree-level matrix elements of
the operators and the effective coefficients $C_{i}^{eff}$, which are
scheme- and (largely) scale-independent:
\begin{equation}
\langle sq'\bar q'\vert {\cal H}_{eff}\vert b\rangle =
\sum_{i,j} C_i^{eff}(\mu)
\langle sq'\bar q'\vert O_j \vert b\rangle ^{\rm tree}  .
\label{me1}
\end{equation}
The effective coefficients multiplying
the matrix elements $< sq'\bar q'\vert O_j^{(q)}\vert b>^{\rm tree}$
may be expressed as \cite{AG97,AKL98-1}:
\bea
C_1^{eff} &=& C_1 + 
\frac{\a_s}{4\p} \, \left( r_V^T +
 \g_{V}^T \log \frac{m_b}{\mu}\right)_{1j} \, C_j  +\cdots ,  
\nonumber \\
C_2^{eff} &=& C_2 +  
\frac{\a_s}{4\p} \, \left( r_V^T +
 \g_V^T \log \frac{m_b}{\mu} \right)_{2j} \, C_j   +\cdots ,
\nonumber \\
C_3^{eff} &=& C_3 -\frac{1}{2N} \frac{\a_s}{4\p} \, (C_t + C_p + C_g)
\nonumber\\
&& + \frac{\a_s}{4\p} \, \left(  r_V^T +
\g_V^T \log \frac{m_b}{\mu}\right)_{3j} \, C_j +\cdots ,  
\nonumber \\
C_4^{eff} &=& C_4 +\frac{1}{2} \frac{\a_s}{4\p} \, (C_t + C_p + C_g)
\nonumber\\
&& + \frac{\a_s}{4\p} \, \left( r_V^T +
\g_V^T \log \frac{m_b}{\mu}\right)_{4j} \, C_j   +\cdots ,
\nonumber \\
C_5^{eff} &=& C_5 -\frac{1}{2N} \frac{\a_s}{4\p} \, (C_t + C_p + C_g)
\nonumber\\
&& +\frac{\a_s}{4\p} \, \left(  r_V^T +
\g_V^T \log \frac{m_b}{\mu}\right)_{5j} \, C_j   +\cdots ,
\nonumber \\
C_6^{eff} &=& C_6 +\frac{1}{2} \frac{\a_s}{4\p} \, (C_t + C_p + C_g)
\nonumber\\
&& +\frac{\a_s}{4\p} \, \left(  r_V^T +
 \g_V^T \log \frac{m_b}{\mu} \right)_{6j} \, C_j   +\cdots ,
\nonumber \\
C_7^{eff} &=& C_7 +\frac{\alpha_{\rm ew}}{8 \, \pi} C_e +\cdots ,
\nonumber \\
C_8^{eff} &=& C_8 \nonumber \\
C_9^{eff} &=& C_9 +\frac{\alpha_{\rm ew}}{8 \, \pi} C_e +\cdots ,
\nonumber \\
C_{10}^{eff} &=& C_{10}~.
\label{EffectiveW} 
\eea
Here, $r_V^T$ and $\gamma_V^T$ are the transpose of the matrices 
derived by Buras {\it et al.} \cite{Burasetal92} and Ciuchini {\it et al.}
\cite{Ciuchinietal95}. The functions $C_t$,
$C_p$ and $C_e$ generate (perturbative) strong interaction phases, essential
for CP violation in $B$ decays \cite{BSS}. They are functions of $k^2$, the
off-shell virtuality in the process $g(k^2) \to q \bar{q}$,
CKM parameters, quark masses and the scale $\mu$ \cite{AG97,AKL98-1}.

A number of remarks on $C_i^{eff}$ is in order. First of all, the scale- and
scheme-dependence in $C_i$ mentioned above are now regulated. However, 
there are still scheme-independent but process-specific terms omitted in
Eq.~(\ref{EffectiveW}) indicated by the ellipses \cite{AG97}. 
The specific constant matrix $r_V^T$ used in our work 
\cite{AG97,AKL98-1,AKL98-2} in defining $C_i^{eff}$ 
has been obtained in the Landau gauge using an off-shell scheme in the
calculation of the virtual corrections \cite{Burasetal92,Ciuchinietal95}.
This raises the spectre of $C_i^{eff}$ becoming gauge dependent \cite{BS98}. 
A remedy of these related problems is a perturbative formulation, in which 
the real and virtual corrections to the matrix elements are calculated in the
NLL approximation in a physical (on-shell) scheme. The 
gauge-dependence in $C_i^{eff}$ will then cancel in 
much the same way as in inclusive decays \cite{Beneke98}. However, for 
exclusive decays, this procedure will bring in a certain cut-off dependence
of $C_i^{eff}$ due to the bremsstrahlung contribution,
for which only a limited part of the phase space can be included in 
$C_i^{eff}$. This sensitivity has to be 
treated as a theoretical systematic error.
Next, as already stated, the coefficients $C_i^{eff}$ are functions of 
$k^2$. In the 
factorization approach, there is no model-independent way to keep track
of this dependence. So, one has to model the $k^2$-dependence or
use data to fix it. At present one varies this parameter in 
some reasonable range \cite{Deshpande90}, 
$\frac{m_b^2}{4}\stackrel{<}{\sim} 
k^2 \stackrel{<}{\sim}\frac{m_b^2}{2}\ $, and
includes this uncertainty in the estimates of the branching ratios.
The $k^2$-related uncertainty in the CP-asymmetries in some cases is 
prohibitively large \cite{kps,AKL98-2}. Clearly, more theoretical work 
and data are needed on these aspects.
%

%%%%%%%%%%%%%%%%%%%%%%%%%%%%%%%%%%%%%%%%%%%%%%%%%%%%%%%%%%%%%%%%%%%%%%%%%%%%
\section{Factorization Ansatz for  $B \to h_1 h_2$}

The factorization Ansatz for the decays $B \to h_1 h_2$ is illustrated 
below on 
the example of the $u$-quark contribution in the operator $O_5$ in
the decay  $B^- \to K^- \omega$ \cite{AG97}, where
 \be
O_5^{(u)} = \left(\bar{s} \g_\mu \, (1-\g_5) b  \right) \,   
                  \left(\bar{u} \g^\mu \, (1+\g_5) u  \right) \quad .
\ee
There are two diagrams which contribute to this decay (see, Fig. 3 in 
ref.~4). Calling the contributions by $D_1$ and $D_2$, the
factorization approximation for $D_1$ is readily obtained:
\bea
D_1 &=& \bra \omega | \bar{u} \g^\mu \, (1+\g_5) \, u|0 \ket \,
\bra K^- | \bar{s} \g_\mu \, (1-\g_5) \, b |B^-\ket \nonumber\\
&=&
 \bra \omega |\bar{u}u_-|0 \ket \, \bra K^- |\bar{s} b_- |B^- \ket 
\quad , 
\eea
where
$\bar{q}q'_- = \bar{q} \, \g_\mu \, (1-\g_5) \, q' $. 
To get $D_2$, one has to write 
the operator $O_5^{(u)}$ in its Fierzed form:
\bea
O_5^{(u)} &=& -2 \, \left( \bar{u}_\b (1-\g_5) b_\a \right)
\left( \bar{s}_\a (1+\g_5) u_\b \right) \nonumber \\
&& =-2 \left[ \frac{1}{N_c} \, \left( \bar{u} (1-\g_5) b \right)
\left( \bar{s} (1+\g_5) u \right) \right. \nonumber\\
&& \left. +
\frac{1}{2}  \, \left( \bar{u} (1-\g_5) \l b \right)
\left( \bar{s} (1+\g_5) \l u \right) \right] \quad , 
\eea
where $\l$ represents the $SU(3)$ colour matrices
and $N_c$ is the number of colours. Now, in the factorization
approximation the second term in the 
square bracket does not contribute and 
one retains only the color-singlet contribution.
This example illustrates the general structure of the matrix elements in
the factorization approach. Thus, generically, one has 
\bea
\bra |h_1 h_2 | {\cal H}_{eff} | B \ket 
&\simeq& \left[ C_{2i-1}^{eff} \bra I \otimes I \ket + \frac{1}{N_c} 
C_{2i}^{eff} \bra I \otimes I \ket \right.\nonumber\\
&&\left. +   C_{2i-1}^{eff} \bra 8 \otimes 8 \ket \right]
+ \left[  C_{2i-1}^{eff} \leftrightarrow C_{2i}^{eff} \right] ~.\nonumber\\
\label{ncxi}
\eea
The factorization approximation amounts to discarding the $\bra 8 \otimes 8 
\ket$ contribution and compensating this by the parameters
$a_{2i}$ and $a_{2i-1}$ ($i=1,...,5$):
\be
a_{2i-1} = C^{eff}_{2i-1} + \frac{1}{N_c} \, C^{eff}_{2i} \quad , \quad
a_{2i} = C^{eff}_{2i} + \frac{1}{N_c} \, C^{eff}_{2i-1} \quad .
\ee
These phenomenological parameters have to be 
determined by experiment. A particularly simple parametrization is obtained 
by replacing $1/N_c$ in Eq.~(\ref{ncxi})  
by a  phenomenological parameter $\xi$:
$$  1/N_c \to \xi ~.$$
With this parametrization, a variety of decays such as $B \to 
(J/\psi,\psi^\prime)(K,K^*)$ and
$B \to (D,D^*)(\pi,\rho)$ etc. yield a universal value of $\xi$.
Further, the parameter $a_1$ in these decays comes out close to its 
perturbative value, obtained by setting $N_c=3$, and  the experimental 
phase of $a_2/a_1$ is found to agree with the one based on 
factorization \cite{NS97}. It is, therefore, tempting to extend
this simplest parametrization to all ten parameters $a_i$ 
in the decays $B \to h_1 h_2$. For a different point of view on the 
parametrization of the penguin amplitudes, see the talk by Cheng 
\cite{Cheng98} and the papers by Ciuchini {\it et al.} 
\cite{Martinellifudge}.

   The numerical values of $a_i$ are given in Table 
1 for three representative values of $N_c$ (equivalently $\xi$) for the
quark level transitions $b \to s$ $[\bar{b} \to \bar{s}]$. They
are evaluated for $k^2=m_b^2/2$, $\mu=m_b/2$ with $m_b=4.88$ GeV, 
$\alpha_s(M_Z)=0.118$, $m_t(m_t)=168$ GeV and wherever necessary, the
CKM-Wolfenstein parameters are set to $A=0.81$, $\rho=0.12$ and $\eta=0.34$,
corresponding to the present best fits \cite{aliapctp97}.

\begin{table*}[t]
\begin{center}
\label{table1}
\caption{Effective coefficients $a_i$ for the
$ b\to s$ [ $ \bar b\to \bar s $] transitions; the
entries for $ a_3$,...,$ a_{10}$ are to be multiplied with
$10^{-4}$.}
\begin{tabular}{c|ccc|}
\hline
        & $N_c=2 $& $ N_c=3 $   & $ N_c=\infty $ \\
\hline
$a_1$  & 0.99 [0.99]& 1.05 [1.05] &  1.16 [1.16] \\
$a_2$  &0.25 [0.25]&  0.053 [0.053] & -0.33 [-0.33] \\
$a_3$  &$-$37$-$ 14i [$-$36$-$14i]  & 48 [48] & 218+  29i [215+  29i] \\
 $a_4$  &$-$402$-$72i [$-$395$-$72i]& $-$439$-$77i[$-$431$-$77i]  &
$-$511$-$87i[$-$503$-$87i]\\
$a_5$  &$- $150$-$14i[$-$149$-$14i]& $-$45[$-$45] & 165+ 29i [162+
29i]\\
$a_6$  &$-$ 547$-$72i[$-$541$-$72i]& $-$575$-$77i[$-$568$-$77i] &
$-$630$-$87i[$-$622$-$87i]\\
$a_7$  & 1.3$-$1.3i [1.4$-$1.3i]& 0.5$-$1.3i[0.5$-$1.3i] &
 $-$1.2$-$1.3i [$- $1.1$-$  1.3i]\\
$a_8$  & 4.4$-$0.7i [4.4$-$0.7i]& 4.6$-$0.4i[4.6$-$0.4i] & 5.0[5.0] \\
 $a_9$  &$-$91$-$ 1.3i [$-$91$-$1.3i]& $-$94$-$1.3i[$-$94$-$1.3i] &
$- $101$-$1.3i[$- $101$-$ 1.3i] \\
$a_{10}$&$-$31$-$0.7i [$-$31$-$0.7i]& $-$14$-$0.4i [$-$14$-$0.4i] &
 20  [20] \\
\hline
\end{tabular}
\end{center}
\end{table*}

A number of observations on the entries in Table 1 is in order:
\begin{itemize}
\item Only the coefficients
$a_1, ~a_4, ~a_6$ and
 $a_9$ are stable against $N_c$-variation,
i.e., they are of $O(1)$ as $N_c \to \infty$, with their relative
magnitudes reflecting the SM dynamics (quark masses and mixing angles). The 
rest
$ a_2, ~a_3, ~a_5, ~a_7, ~a_8$ and $a_{10}$ being of $O(1/N_c)$ are unstable 
against the variation of $N_c$.
\item The coefficients $a_1,~a_2,~a_4, ~a_6$ and $a_9$ can be determined
by measuring the ratios of some selected branching ratios. This has been 
studied extensively in the paper with Kramer and L\"u \cite{AKL98-1},
where detailed formulae and their (reasonably accurate) approximate forms
are given. These ratios will be helpful in testing the predictions of
the factorization approach in forthcoming experiments.
\item The QCD-penguin coefficient $a_3$ and the electroweak-penguin
coefficients
$a_7, ~a_8$ and $a_{10}$ are numerically very small. Hence, it will be
difficult to measure them.
 \end{itemize}
\subsection{Classification of Factorized Amplitudes}
 In the context of Tree-decays, a classification of
the two-body decay amplitudes was introduced by Stech and co-workers 
\cite{BSW87}.
These classes, concentrating now on the $B \to h_1 h_2$ decays, are the
following:
\begin{itemize}
\item Class-I, involving decays in which only a charged 
meson can be generated directly from a colour-singlet current, as in  
$B^0 \to \pi^+ \pi^-$. For this class, ${\cal M}(B \to h_1 
h_2) \propto$ $a_1$. 
\item Class-II, involving decays in which the meson generated 
from the weak current is a neutral meson, like in $B^0 \to \pi^0 \pi^0$.
For this class, ${\cal M}(B \to h_1 h_2) \propto$ $a_2$.
\item Class-III, involving the interference of class-I and class-II decay 
amplitudes, ${\cal M}(B \to h_1 h_2) \propto$ $a_1 + r a_2$, where 
$r$ is a process-dependent (but calculable in terms of form factors etc.)
constant.
Some examples are $ B^\pm \to \pi^\pm \pi^0$, $ B^\pm \to \pi^\pm 
\rho^0$ and $ B^\pm \to \pi^\pm \omega$.
\end{itemize}
This classification has been extended to the decays involving penguin 
operators \cite{AKL98-1}. In the $B \to h_1 h_2$ decays, one now has two
additional classes:
\begin{itemize}
\item Class-IV, involving decays whose amplitudes contain one (or more) 
of the dominant penguin coefficients $a_4$, $a_6$ and $a_9$, with 
constructive interference among them. Their amplitudes have the generic form:
\begin{eqnarray}
{\cal M}( B^0 \to h_1^\pm h_2^\mp) &\simeq&\alpha_1 a_1  +
\sum_{i=4,6,9}\alpha_i a_i~ +..., \label{classiv}
\\ \nonumber
{\cal M}( B^0 \to h_1^0 h_2^0) &\simeq&\alpha_2 a_2 + \sum_{i=4,6,9}\alpha_i
a_i~ +..., \\ \nonumber
{\cal M}( B^\pm \to h_1^\pm h_2^0) &\simeq&\alpha_1 (a_1 + r a_2) +
\sum_{i=4,6,9}\alpha_i a_i~ +...,
\end{eqnarray}
with the second (penguin-induced) term
dominant in each of the three amplitudes. The ellipses indicate possible
contributions from
the coefficients $a_3,a_5,a_7,a_8$ and $a_{10}$ which can be neglected
for this class of decays.
The coefficients $\alpha_j$ are process-dependent and  contain the CKM matrix
elements, form factors etc.
\end{itemize}
 Examples of Class-IV decays are quite abundant. They  include decays
such as $B^\pm \to K^\pm \pi^0$, $B^\pm \to K^\pm \eta^{(\prime)}$, which
involve
$a_1 + r a_2$ as the tree amplitude, and ${B}^0 \to {K}^0 \pi^0$,
${B}^0 \to {K}^0 \eta^{(\prime)}$ (and charged conjugates), which 
involve $a_2$ from
the tree amplitude. Finally, the pure-penguin decays, such as $B^\pm \to
\pi^\pm K^0$, $B^\pm \to K^\pm \bar{K}^0$ etc. 
naturally belong here. Several of these decays have been measured by the
CLEO collaboration \cite{cleo}.
\begin{itemize}
\item Class-V, involving decays with strong $N_c$-dependent penguin 
coefficients $a_3$, $a_5$, $a_7$ and $a_{10}$, interfering significantly 
with one of the dominant penguin coefficients.  
Decays in which the dominant penguin coefficients interfere destructively
are also included here. 
\end{itemize}
Examples of this class are:
$B^\pm \to \pi^\pm \phi$, ${B}^0 \to \pi^0 \phi$,
${B}^0 \to \eta^{(\prime)} \phi$.
 In all these cases, the amplitudes are proportional to
the linear combination $[a_3 + a_5 -1/2 (a_7 + a_9)]$.
Examples of decays whose amplitudes are proportional to the dominant penguin
coefficients interfering destructively are: $B^\pm \to K^\pm \phi$, $B^0 \to
K^0 \phi$. The above five classes exhaust all cases.

 One expects that only  
class-I and class-IV decays (and possibly some class-III decays) can be
predicted with some reasonable theoretical accuracy (typically a factor 2).
In almost all of the decays studied in these classes \cite{AG97,AKL98-1}, 
the variation
of the branching ratios with $N_c$ is not very marked. Hence, the effective
coefficients extracted from data should come out rather close to
their perturbative QCD values.
Decays in other classes involve, in most cases,
large and delicate cancellations, and hence their decay rates 
\cite{AKL98-1} (and CP asymmetries) are difficult to be predicted 
reliably \cite{AKL98-2}. An example is the decay $B^\pm \to \omega K^\pm$,
which has been measured by the CLEO collaboration \cite{cleo} but whose
decay rate varies by more than an order of magnitude in the range $0 < \xi < 
0.5$ in the present approach \cite{AG97,AKL98-1}.

\section{Charm Content of the $\eta$ and $\eta^\prime$ and the Role of QCD
Anomaly in the Decays $B\to K \eta^{(\prime)},K^{*} 
\eta^{(\prime)}$}
  Before comparing this framework with data, we discuss the decays
$B \to K \eta^\prime (\eta)$ and $B \to K^*\eta^\prime (\eta)$,
which have received lot of interest lately \cite{Jimreview}.
To be specific, we concentrate on the decay $B^\pm \to K^\pm \eta^\prime 
(\eta)$.

 In the factorization approach, the matrix element for the decay $B^\pm
\to K^\pm \eta^\prime$ can be expressed as
\be
{\cal M} = - \frac{G_F}{\sqrt{2}} \, V_{cb} \, V_{cs}^* \, a_2 \,
\bra \eta'(q)|\bar{c} \g_\mu \g_5 c |0 \ket \, \bra
K(p')|\bar{s} \g^\mu b |B(p) \ket ~.
\ee
Defining 
\be
\bra \eta^\prime(q)|\bar{c} \g_\mu \g_5 c |0\ket \equiv -i 
f^{(c)}_{\eta^\prime} \, q_\mu ~,
\ee
the quantities $f^{(c)}_{\eta^\prime}$ and $f^{(c)}_{\eta}$
measure the charm content of the $\eta^\prime$ and $\eta$, 
respectively. Using this notation, the coupling constant $f_{\eta_c}$,
defined analogously,
\be
\bra \eta_c(q)|\bar{c} \g_\mu \g_5 c |0\ket \equiv -i f_{\eta_c} \, q_\mu ~,
\ee
can be used to normalize them. 
 The constants $f^{(c)}_{\eta^\prime}$ and $f^{(c)}_{\eta}$
have been determined in a variety of ways. Here the following two methods
are reviewed \cite{AG97,ACGK97}.
\begin{itemize}
\item $f_{\eta^\prime}^{(c)}$ and $f_{\eta}^{(c)}$ via the 
$\eta$-$\eta^\prime$-$\eta_c$ mixing \cite{AG97}
\end{itemize}
This is a purely phenomenological approach. 
 One admits a small $|c \bar{c}\ket$ admixture in the 
$SU(3)$-singlet state vector $|\eta_0 \ket$,
characterized by  $\theta_{cc}$.
In the small-$\tan \theta_{cc}$ limit, and using one-mixing-angle 
($\theta$) formalism for the $(\eta$-$\eta^\prime)$ complex, one 
can write down the following relations \cite{AG97}:  
\bea
\label{relation}
f^{(c)}_{\eta^\prime} &\simeq&  \cos \theta \, \tan \theta_{cc} \, 
f_{\eta_c} ~, \nonumber\\
f^{(c)}_{\eta} &\simeq&  \sin \theta \, \tan \theta_{cc} \, f_{\eta_c} ~.
\eea
Using the observed decay width \cite{PDG96}
\be
\G(\eta_c \to \g\g) = \frac{4(4\p\a)^2 \,
f^2_{\eta_c}}{81 \p m_{\eta_c}} = 7.5^{+1.6}_{-1.4} \ \mbox{KeV} ~,
\ee
one obtains $f_{\eta_c} = 411 $ MeV from the central value. The 
mixing angle $\theta_{cc}$ can be determined from the 
ratio of the following radiative $J/\psi$-decays \cite{PDG96}:
\bea
\frac{{\cal B}(J/\psi \to \eta_c \g)}{{\cal B}(J/\psi \to \eta' \g)} &=&
\frac{(1.3 \pm 0.4) \times 10^{-2}}{(4.31 \pm 0.30) \times 10^{-3}}
\nonumber\\
 &\simeq &
\left(\frac{k_{\eta_c}}{k_{\eta'}}\right)^3 \,
\frac{1}{\cos^2 \theta \, \tan^2 \theta_{cc}} \quad .
\eea
This leads to a value $|\theta_{cc}|\simeq 0.014$, yielding \cite{AG97}:
\bea 
|f^{(c)}_{\eta^\prime}| &=&| \cos \theta \, \tan
\theta_{cc} \, f_{\eta_c}| \simeq 5.8 \, \mbox{MeV} ~,
\nonumber\\
|f_\eta^{(c)}| &=& |\sin \theta \, \tan \theta_{cc} \, f_{\eta_c}| \simeq
2.3 \, \mbox{MeV} ~.
\quad
\eea
Note that the signs of $f^{(c)}_{\eta^\prime}$ and $f_\eta^{(c)}$ are
not determined in this method. In the two-angle mixing formalism for the
$(\eta$-$\eta^\prime)$ complex \cite{leutwyler}, the angle $\theta$
in the expressions for  $ f^{(c)}_{\eta^\prime}$ and 
$f_{\eta_c}$ gets replaced by  $\theta_0$, the angle in the singlet sector.
Since $|\theta_0| < |\theta|$ (typical values are: $\theta \simeq 
-22^\circ$ and $\theta_0=-(4 - 9)^\circ$), the value of 
$f_{\eta_c}$ is reduced, yielding $|f_\eta^{(c)}| \simeq 1$ MeV. 
 \begin{itemize}
\item $f_{\eta^\prime}^{(c)}$ and $f_{\eta}^{(c)}$ via QCD 
Anomaly \cite{ACGK97}
\end{itemize}
In this case, the matrix elements are modeled by annihilating the
charm-anticharm quark pair into two gluons, effecting the decay $b \to s g 
g$, followed by the transition $gg \to \eta^{(\prime)}$ (see Fig.~1 in
ref.~5).
The first part of this two-step process, i.e., $b \to s gg$, has been worked 
out by Simma and 
Wyler \cite{SW90}, and their result can be transformed in the language of the
effective theory:
\be
  \label{newop}
H_{\mathrm{eff}}^{gg}=- \frac{\alpha_s}{2\pi} a_2
\frac{G_F}{\sqrt{2}} V_{cb} V_{cs}^* \Delta i_5
\bigl(\frac{q^2}{m_c^2}\bigr) O_{sgg}~,
\ee
where the non-local operator $O_{sgg}$ is given by
\be
O_{sgg}=\frac{1}{k_1 \cdot k_2} G^{\alpha \beta}_a (D_{\beta}
\tilde{G}_{\alpha \mu} )_a \, \overline{s}\gamma^{\mu} (1-\gamma_5) b
~,
\ee
with $\tilde{G}_{\mu \nu}=\frac{1}{2} \epsilon_{\mu \nu \a \b} G^{\a
\b}$ and $q^2 = (k_1 + k_2)^2 = 2k_1 \cdot k_2$, where $k_1$ and $k_2$ are
the momenta of the two gluons.  The function $\Delta i_5 (z)$ is given by
\be
  \label{deltai5}
  \Delta i_5 (z) = -1 +\frac{1}{z} \Bigl[ \pi -2 \tan^{-1} (\frac{4}{z}
  -1)^{1/2} \Bigr]^2, \ \mbox{for $0<z<4$} ~.
\ee
One can expand this function in $q^2/m_c^2$, which makes it clear that
the leading term in $H_{eff}^{gg}$ induces a power ($1/m_c^2$) 
correction. In fact, in this form $H_{eff}^{gg}$ becomes the
chromo-magnetic analogue of the corresponding operator in the decay
$B \to X_s \gamma$ discussed by Voloshin \cite{Voloshinbsg}.
Then, on using the equation of motion and the $U_1$ axial-anomaly: 
\be
\langle \eta^{(')} | \frac{\alpha_s}{4\pi} G^{\alpha \beta}_a
\tilde{G}_{\alpha \beta,a} |0 \rangle = m_{\eta^{(')}}^2
f_{\eta^{(')}}^u ~,
\ee
one gets \cite{ACGK97}:
\be
f^{(c)}_{\eta'} \simeq - 3.1 [-2.3]\, ~\mbox{MeV} ~,
\ee
\be
f_\eta^{(c)}  \simeq -1.2  [-0.9]\, ~\mbox{MeV}
~,
\ee
corresponding to the value $m_c=1.3$ GeV $[m_c=1.5$ GeV].
 The two calculations give (within a factor 2) consistent results, with
the anomaly method determining both the magnitudes and signs. The
charm contents of the $\eta^\prime$ and $\eta$  are,
however, found to be small in both approaches.
\section{Comparison with Data and Outlook}
In Table 2, the branching ratios, averaged over the charge-conjugated 
modes, for the decays $B \to PP$ involving two pseudoscalar mesons 
are shown. The entries in this table \cite{AKL98-1} have been
calculated using the BSW-model \cite{BSW87} and [Lattice QCD/QCD sum rule]
form factors. Experimental numbers from CLEO \cite{cleo} are 
shown in the last column.
% Branching ratios for the other decays $B \to PV$ and $B \to 
%VV$, involving the pseudoscalar-vector and vector-vector mesons, can be 
%seen 
%elsewhere \cite{AG97,AKL98-1}. Estimates of the CP asymmetries in these
%decays are given in the paper with Kramer and L\"u \cite{AKL98-2}.

We conclude this contribution with a number of remarks.

\begin{itemize}
\item All five decays measured by the CLEO collaboration shown in 
Table 2 are penguin-dominated class-IV decays. The estimates based on the
factorization model are in reasonable agreement with data, except perhaps for
the decay $B^+ \to K^+ \eta^\prime$ for which experiment lies 
(approximately) a factor 2 higher. All upper limits are in accord with the
estimates given here. Thus, QCD-penguins in $B$ decays are measurably large
but not anomalous.
\item It is fair to conclude that the QCD-improved factorization 
framework discussed here provides a
first step towards  understanding exclusive two-body $B$ decays.
However, there are many open theoretical questions and more work is needed.
In particular, most class-V decays is a hrad nut to crack.
 \item Finally, non-leptonic $B$ decays provide new avenues to determine the
CKM parameters. At present no quantitative conclusions can be drawn as
the experimental errors are large, but potentially some of these decays  
will provide complementary information on the CKM 
parameters \cite{fleischer} to the one from the unitarity constraints.
\end{itemize}

\begin{table*}[t]
\begin{center}
\label{table2}
\caption{ $B\to PP$ Branching Ratios (in units of $10^{-6}$). } 
\begin{tabular} {|l|c|c|c|c|c|}
\hline
Channel &  Class & $N_c=2$ &  $N_c=3$ & $N_c=\infty$ & Exp. \\
\hline
$B^0 \to \pi^+ \pi^-$ & I & $9.0 \;[11 ]$ &$10.0 \;[12 ]$ 
 & $12 \;[15] $ & $ < 15$ \\
$B^0 \to \pi^0 \pi^0$  & II &  $0.35 \;[0.42]$ &$0.12 \;[0.14]$ 
 &$0.63 \;[0.75]$ &$<9.3$ \\
$ B^0 \to \eta^\prime \eta^\prime$  & II & $0.05 \;[0.07]$ &$0.02 \;[0.02]$
 &$0.09 \;[0.10] $ & $<47$\\
$ B^0 \to \eta \eta^\prime$  & II & $0.19 \;[0.22]$ &$ 0.08 \;[0.10] $ 
 &$0.29 \;[0.34]$ & $<27$\\
$ B^0 \to \eta \eta$  & II & $0.17 \;[0.20]$ &$0.10 \;[0.11] $ 
&$0.24 \;[0.29]$ & $<18$\\
$ B^+ \to \pi^+ \pi^0$ & III & $6.8 \;[8.1 ]$ &$5.4 \;[6.4 ]$
 &$3.0 \;[3.6 ]$ & $ < 20$\\
$ B^+ \to \pi^+ \eta^\prime$  & III & $2.7 \;[3.2 ]$ &$2.1 \;[2.5 ]$
 &$1.1 \;[1.4]$ & $<12$\\
$ B^+ \to \pi^+ \eta$  & III & $3.9 \;[4.7 ]$ &$3.1 \;[3.7 ]$
 &$1.9 \;[2.2 ]$ & $<15$\\
$ B^0 \to \pi^0 \eta^\prime$  & V & $0.06 \;[0.07] $ &$0.07 \;[0.09] $ 
 &$0.11 \;[0.13] $ & $<11$\\
$ B^0 \to \pi^0 \eta$ & V  & $0.20 \;[0.24] $ &$0.23 \;[0.27]$ 
 &$0.30 \;[0.36]$ & $<8 $\\

$ B^+ \to K^+ \pi^0$  &  IV & $9.4 \;[11 ]$  & $10 \;[12]$  &
  $12 \;[15]$  &15 $ \pm 4\pm 3 $ \\
$B^0 \to K^+ \pi^-$  & IV & $14 \;[16]$ &$15 \;[18]$ &
$18 \;[21]$ &$14 \pm 3 \pm 2$\\
$B^0 \to K^0 \pi^0$  & IV & $5.0 \;[5.9 ]$  & $5.7 \;[6.8 ]$  &
  $7.4 \;[8.9 ]$  &$ <41$\\
$ B^+ \to  K^+ \eta^\prime$  & IV & $21 \;[25]$  & $25 \;[29]$  &
  $35 \;[41]$  &$ 74^{+8}_{-13}\pm 9 $ \\
$B^0 \to K^0 \eta'$  & IV &  $20 \;[24]$  & $25 \;[29]$  &
  $35 \;[41]$  &$ 59^{+18}_{-16}\pm 9 $\\
$ B^+ \to  K^+ \eta$  & IV & $2.0 \;[2.3 ]$  & $2.4 \;[2.7 ]$  &
  $3.4 \;[3.9 ]$  &$ <14$\\
$B^0 \to K^0 \eta$  & IV & $1.7 \;[1.9 ]$  & $2.0 \;[2.2 ]$  &
  $2.6 \;[3.0 ]$  &$ <33 $\\
$ B^+ \to \pi^+ K^0$  & IV & $14 \;[17]$ & $16 \;[20]$ 
& $22 \;[26]$ & $14 \pm 5 \pm 2 $\\
$ B^+ \to K^+ \bar K^0$  & IV & $0.82 \;[0.95]$ & $0.96 \;[1.1 ]$
& $1.3 \;[1.5 ]$ & $ <21$\\
$ B^0 \to K^0\bar K^0$ & IV  &  $0.79 \;[0.92]$ &$0.92 \;[1.1]$
& $1.2 \;[1.4 ]$ & $ <17$\\
\hline
\end{tabular}\end{center}
\end{table*}

\end{document}